\documentclass[twoside,11pt]{article}

\usepackage{jmlr2e}
\usepackage{amsmath}
\usepackage{algorithm}
\usepackage[noend]{algpseudocode}

\jmlrheading{1}{2015}{1-48}{1/15}{1/15}{Ashish Kapoor, Paxon Frady, Stefanie Jegelka, William B. Kristan and Eric Horvitz}
\ShortHeadings{Inferring and Learning from Neuronal Correspondences}{Kapoor, Frady, Jegelka, Kristand and Horvitz}
\firstpageno{1}

\begin{document}

\title{Inferring and Learning from Neuronal Correspondences}

\author{\name Ashish Kapoor \email akapoor@microsoft.com \\
       \addr Microsoft Research\\
       One Microsoft Way\\
       Redmond, WA 98052, USA
       \AND
       \name E. Paxon Frady \email efrady@ucsd.edu \\
       \addr Department of Neurobiology\\
       University of California\\
       San Diego, CA 92093, USA
       \AND
       \name Stefanie Jegelka \email stefje@mit.edu\\
       \addr Department of Electrical Engineering and Computer Science\\
       Massachusetts Institute of Technology\\
       Cambridge, MA 02139, USA
       \AND
       \name William B. Kristan \email kristan@ucsd.edu \\
       \addr Department of Neurobiology\\
       University of California\\
       San Diego, CA 92093, USA
       \AND
       \name Eric Horvitz \email horvitz@microsoft.com \\
       \addr Microsoft Research\\
       One Microsoft Way\\
       Redmond, WA 98052, USA
       }

\editor{TBD}

\maketitle

\begin{abstract}
We introduce and study methods for inferring and learning from correspondences among neurons. The approach enables alignment of data  from distinct multiunit studies of nervous systems. We show that the methods for inferring correspondences combine data effectively from cross-animal studies to make joint inferences about behavioral decision making that are not possible with the data from a single animal. We focus on data collection, machine learning, and prediction in the representative and long-studied invertebrate nervous system of the European medicinal leech.  Acknowledging the computational intractability of the general problem of identifying correspondences among neurons, we introduce efficient computational procedures for matching neurons across animals.  The methods include techniques that adjust for missing cells or additional cells in the different data sets that may reflect biological or experimental variation. The methods highlight the value harnessing inference and learning in new kinds of computational microscopes for multiunit neurobiological studies.
 \end{abstract}


\section{Introduction}

Neurobiologists have long pursued an understanding of the emergent phenomena of nervous systems, such as the neuronal basis for choice and behavior. Much research on neuronal systems grapples with the complex  dynamics of interactions among multiple neurons. New techniques, such as calcium imaging, voltage-sensitive dye (VSD) imaging \citep{cacciatore,gonzalez} and multi-unit electrode recordings, enable larger views of nervous systems. However, for many experimental preparations, the amount of data that can be collected via tedious experiments is limited. As an example, data from voltage-sensitive dyes are time-limited because of bleaching of the dyes and also neuronal damage caused by phototoxicity.

We have developed methods for combining the data from multiple experiments to pool data on neural function. The approach allows us to make inferences from data sets that are impossible to obtain from individual preparations. Coalescing the data from multiple experiments is an intrinsically difficult problem because of the difficulty in matching cells and their roles across animals.  Variation is observed in nervous systems of individual animals based on developmental differences as well as artifacts introduced in the preparation and execution of experiments.  Developing a means for identifying correspondences in cells across animals would allow data to be pooled from multiple animals supporting deeper inferences about neuronal circuits and behaviors.

We focus specifically on experimental studies of neurons composing the ganglia of H. verbana \citep{briggman}.   The leech has a stereotypical nervous system consisting of repeating packets of about 400 neurons. About a third of these neurons have been identified, and these neurons can be found reliably in different animals. The remaining two-thirds of neurons have yet to be identified, but are believed to maintain similar properties and functional roles across animals. The general problem of correspondence matching of the cells of two different animals is illustrated in Figure 1. We seek to identify neurons that are equivalent across ganglia obtained from different animals. For example, the red cell in animal $a$ has several candidate correspondences in animal $b$, but with varying degrees of similarity (indicated by shades of red). Ideally, we will find a one-to-one match via jointly considering multiple similarities. With only two animals, this problem can be mapped to a bipartite graph-matching task and can be solved optimally \citep{munkres}.  However, we want to jointly solve the matching task for larger numbers of animals. Such matching across multiple graphs defined by the individual nervous systems is an intractable problem in the NP-hard class \citep{papa}. In addition, the problem is even more difficult because such matching must also take into account variations in the numbers and properties of neurons observed in different animals. These variations can be due to both developmental differences (e.g., some neurons may be missing or duplicated) and experimental artifacts (e.g., some neurons may be out of the plane of focus or destroyed in the delicate dissection).  A key challenge in this endeavor is the formulation of a similarity measure that takes into account physical parameters of cells, such as their size and location, as well as their functional properties.

\begin{figure}
\includegraphics[width=\textwidth]{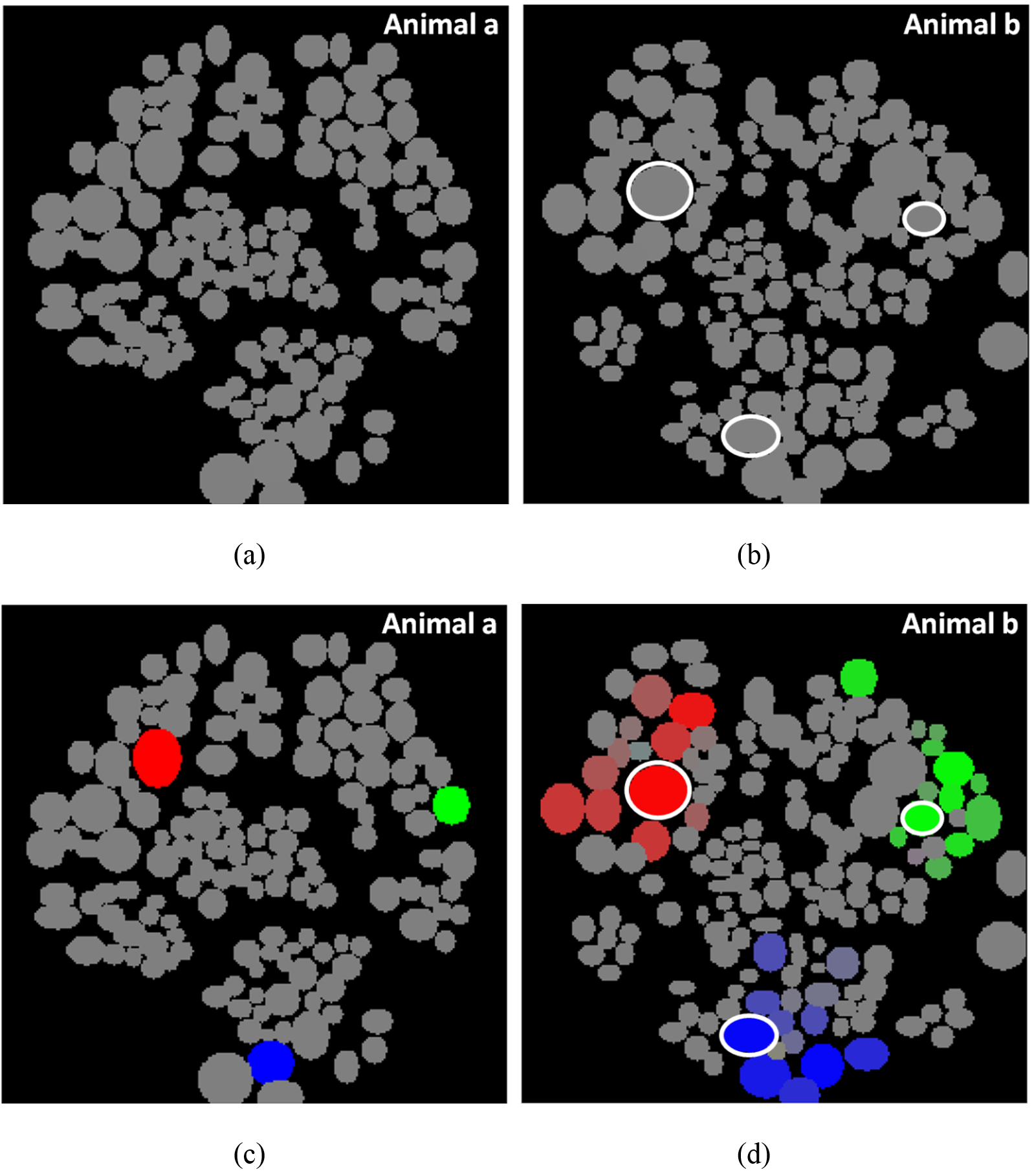}
\caption{Challenge of identifying correspondence among neurons in ganglia from different animals.  Given compatibility constraints, a correspondence algorithm seeks a one-to-one mapping between neurons in two H. verbana preparations. The goal is to find out correspondences between cells in each of the animals. The color coding illustrates compatibility constraints: Feasible matches for the highlighted red, green, and blue cells in animal 1 (c) are found in animal 2 (d), as indicated with matched colors. The degree of feasibility of the matches is depicted via shading of cells in animal $b$, where the most compatible cell for each source neuron in animal $a$ is highlighted with a white border. Although the figure shows only two animals, such compatibility constraints occur across all pairs of the 6 animals used.}
\end{figure}

\section{Machine Learning Framework}
\begin{table}
\caption{The proposed algorithmic framework}\label{algo}
\begin{center}
\begin{tabular}{l l l}
\hline
\hline
        &                             & Given training data with pairs of match and\\
        &                             & non-match cells, estimate the parameter matrix\\
Step 1: & Learn Compatibility Measure & $A$ that defines the compatibility function\\
        &                             & $f^{ab}(i,j)$, between any  $i^{th}$ and  $j^{th}$ neurons for\\
        &                             & all animal pairs. \\
\hline
        &                             & Start with an initialized empty match set $\mathcal{S}_0$.\\
        &                             & Iteratively determine the next best match\\
Step 2: & Recover Correspondence Map  & $\mathcal{M}_t$ by solving equation \ref{equ:match}\\
        &                             & and update $\mathcal{S}_t = \mathcal{S}_{t-1} \cup \mathcal{M}_t$.\\
        &                             & End when all the cells are matched. \\
\hline
        &                             & Construct the matrix $Y$ that aggregates\\
        &                             & data from all the animals, where each \\
Step 3: & Infer Missing Data          & row corresponds to cells and are permuted\\
        &                             & according to the matching. Use Probabilistic \\
        &                             & PCA on $Y$ to infer the missing data.\\
\hline
\hline
\end{tabular}
\end{center}
\end{table}

We use a set of neuronal data collected in \cite{briggman}, which consists of optical VSD recordings from populations of $123$-$148$ neurons in a mid-body segmental ganglion from six different leeches. Earlier research on this data identified neurons involved in decision-making.  In particular, the study aimed at understanding the roles of neuronal populations in decisions to swim or to crawl following stimulation.  Sensory neurons (DP nerve) were stimulated in such a way that would elicit, with equal ($0.5$) likelihood, swimming or crawling. This previous study considered single cell activations and joint analysis of neurons using dimensionally reduction techniques of PCA and LDA. However, these techniques were limited to one animal at a time.  In the current study, we propose a framework that analyzes data across-animals to increase the power of the analysis.

Rather than using a handcrafted measure, we have employed a machine-learning framework that relies on supervised training data. This algorithm estimates an appropriate similarity function between neurons in different animals based on a training set of high-confidence correspondences. These correspondences are readily identified neurons in the nervous systems of H. verbana (Muller et al., 1981). An important capability of the algorithm is to take into account the probabilistic nature of inferred correspondences. The algorithm begins by learning a weighting function of relevant features that maximizes the likelihood of matches within the training set. The next step of the approach is to jointly solve the correspondence matching problem for neurons across animals, while considering potential missing or extra cells in each animal. The final step is to consider correspondences with functions that fill in missing data. As we will demonstrate below, pooling neurophysiological data from multiple studies in a principled manner leads to larger effective data with greater statistical power than the individual studies.

Specifically, the pipeline for the methodology includes three steps (detailed in Table \ref{algo}): (1) Determining a similarity score across pairs of cells, (2) recovering correspondences that are consistent with the similarity measure, and (3) estimating missing data. We describe these steps in detail below:

\subsection{Learning Similarity Measure for Cells}
The goal in Step 1 is to learn a similarity function $f^{ab}(i,j)$ indicating the feasibility of a match between the $i^{th}$ cell in animal $a$ and the $j^{th}$  cell in animal $b$. The most desirable characteristic for such a function is a high positive value for likely matches and diminishing values for poor matches. Such a characteristic is captured by the exponentiation of a negative distance measure among sets of features that represent multiple properties of cells. Formally:

\begin{equation}
f^{ab}(i,j)=e^{-[\phi(i,a)-\phi(j,b)]^T A[\phi(i,a)-\phi(j,b)]}
\end{equation}

Here, $\phi(\cdot)$ are $d$ dimensional feature representations for the individual cells for each animal and summarizes physical (e.g.,  size, location etc.) and functional (e.g., optical recordings) properties. $A$ is a $d \times d$ parameter matrix with positive entries that are learned from data. Intuitively, the negative log of the similarity function is a distance function between the feature representations: a zero distance between two feature vectors result in highest similarity measure of 1, whereas representations at further distance away in the feature space have a diminishing value. The matrix $A$ parameterizes this distance measure. Given training data consisting of several probable pairs of matched neurons, we use $A$ to solve an optimization problem. We describe the details below.

The following list of features were used in our work:
\begin{itemize}
\item{{\bf Structural features:} Absolute position of the cell with respect to the entire observed frame, relative position of cell in relation to the entire ganglion, absolute size of the cell in pixels, indicator vector specifying packet the neuron is located in (among Central, Left Anterior, Left Posterior, Right Anterior, Right Posterior or Central Posterior packet), and relative position coordinate of the neuron in its respective packet.}
\item{{\bf Functional features:} Coherence of electrophysiological observations with swim oscillations and single cell discrimination time  (see \cite{briggman}) that distinguishes from swim versus the crawl behavior.}
\end{itemize}

Intuitively, the optimization problem finds the parameter $A$ that minimizes the distance between pairs of cells that were tagged as matches, while maximizing the distance among other pairs. Formally, parameter $A$ of the compatibility measure is estimated by minimizing the objective:
\begin{equation}
A^* = \arg \min_{A} â¡\sum_{i,j}{[-2 \logâ¡ f^{ab}(i,j) + \log \sum_{j'\in b, j' \neq j} f^{ab} (i,j') + \log \sum_{i' \in a, i' \neq i} f^{ab}(i',j)]}
\end{equation}
subject to the constraint that all entries of $A$ are positive. The sum is over all the labeled training pairs $(i,j)$ tagged as likely matches. Intuitively the first term $-2 \logâ¡ f^{ab} (i,j)$ in the objective prefers solutions that would collapse the distance between matched pairs to zero, while the rest of the terms prefer a solution where the distance between the rest of the cells are maximized. The optimization is straightforward and a simple gradient descent will always find a locally optimal solution. Note that a more appropriate constraint is positive semi-definite condition on $A$, however we suggest using a non-negativity constraint due to simplicity in optimization with almost no reduction in performance of the pipeline.

\subsection{Correspondence Matching}
In a second step, we calculate the correspondence matches across all the animals. Instead of calculating all the matches simultaneously, the framework follows an iterative procedure: future matches are made not only by using the similarity function, but also by comparing the geometric and structural relationship of the candidates to the past matches. Besides considering the distances induced by the similarity function (i.e. $-\log â¡f^{ab}(i,j)$), and unlike past work on graph matching \citep{williams,bunke}, the proposed method utilizes knowledge of â€œlandmarksâ€ by inducing constraints that impose topological and geometric invariants. This match-making algorithm considers the iteration $t$ and denotes the set of already determined matches by $\mathcal{S}_t$. The algorithm then determines $\mathcal{M}^{t+1}$, the next set of neurons from all the animals to be matched by solving the following optimization task:
\begin{equation}
\mathcal{M}^{t+1} = \arg \min_{\mathcal{M}} â¡\sum_{\mbox{all pairs} (i,j) \in \mathcal{M}} -\logâ¡ f^{ab}(i,j)+ \lambda D_{LM}(i, j, \mathcal{S}_t)
\label{equ:match}
\end{equation}
Here $\lambda$ is the trade-off parameter that balances the compatibility measure with landmark distances $D_{LM}(\cdot)$  from the matches recovered in all the prior iterations. The landmark distance computation provides important structural and topological constraints for solving the correspondence tasks. Given anchor points, the landmark distances attempt to capture structural and locational relationship with respect to the available landmarks. There are several options such as commute distance \citep{mckay,lovasz} on a nearest-neighbor graph, or Euclidean distance computed by considering either the locations or the feature representation of the neurons. In our experiments, we compute landmark distances between neuron $i$  in animal $a$  and neuron $j$ in animal $b$ with respect to a set of anchor points S as:
\begin{equation}
D_{LM}(i,j,S)=  \sum_{i' \in \mathcal{S}} \logâ¡ f^{aa} (i,i') - \sum_{j' \in \mathcal{S}} \logâ¡ f^{bb} (j,j').
\end{equation}
The optimization problem in the above equation is solved using off-the-shelf energy minimization procedures \citep{boykov,minka}. The set of the newly discovered matches are then included and the process is repeated until all matches stay the same (settle). Essentially, the goal is to find a set of matched neurons across all the animals such that objective function is minimized. We start with a reasonable initialization of solution (for example by solving for consecutive pairs of animals). This solution is iteratively refined by considering data drawn from one study at a time and searching for a replacement neuron which would lower the total energy. Such replacements continue until no further minimization is observed.

Utilizing landmarks are appropriate as an informative signal for matching neurons in the leech, because there is a typified geometric structure. Although soma positions do vary from animal to animal, often certain somas remain arranged with particular geometrical relationships. For instance the Nut and AE cells typically form a box-like pattern, the N and T sensory neurons usually will be arranged in a hemi-circle along the packet edge, which often will wrap around the AP cell. These types of arrangements are useful for identification of cells by eye, and we extend our algorithm to utilize these relationships.

The framework is extended to handle poor matches and missing cells by considering a {\em sink} cell in every animal. The sink cell has a fixed cost of matching, denoted as c, and acts as a threshold such that neuron matches with costs greater than c are disallowed. The sink cells are a soft representation of the probability that a particular neuron was not visible during a given preparation.
\begin{figure}[t]
\includegraphics[width=\textwidth]{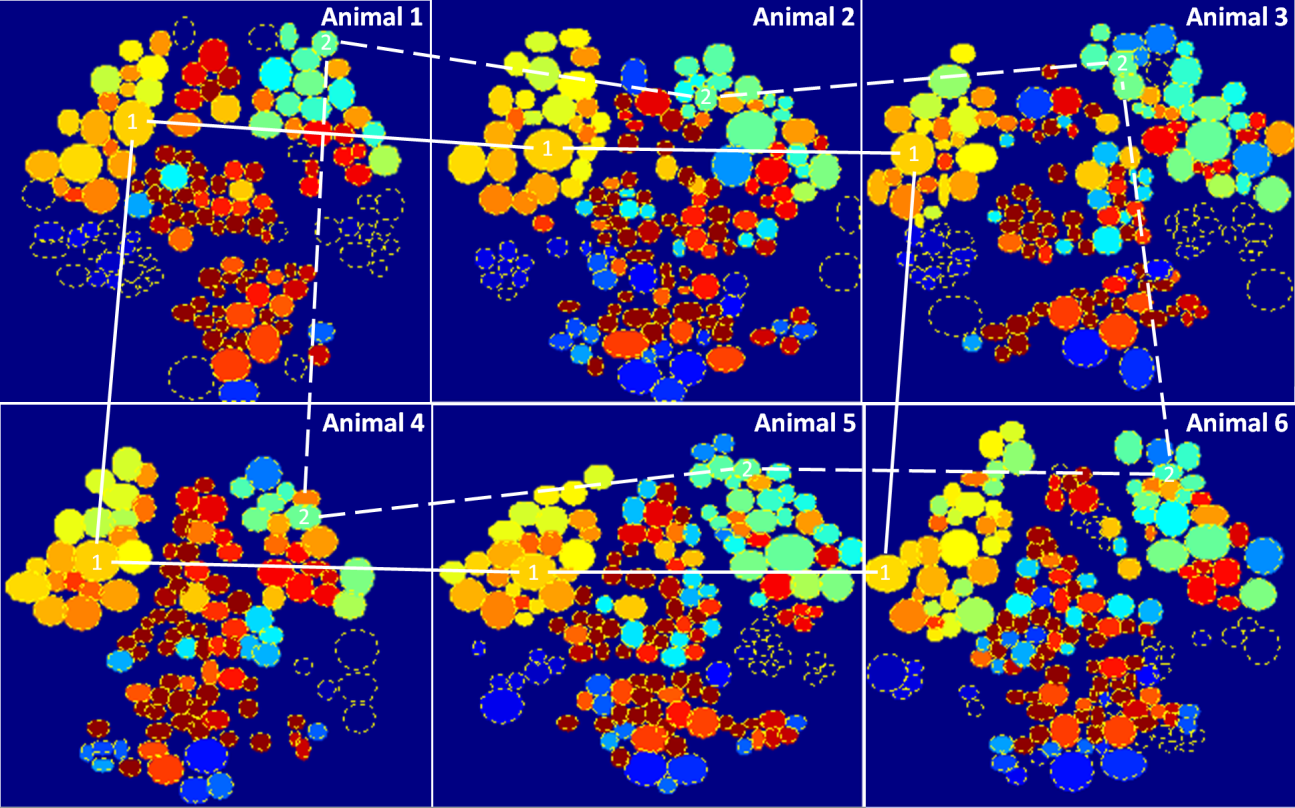}
\caption{Cell correspondences inferred across six H. verbana. Graphics show results of the correspondence matching procedure across six animals. Color coding indicates the correspondences, where matched cells across different animals share the same color. We highlight two cells (depicted as 1 and 2) and show the matches as lines linking neurons across the animals. Several cells remain unmatched and are depicted using the dashed lines (unfilled interior). The algorithm is capable of handling partial matches where cells are not present in all the six animals due to true structural differences or losses either in their preparation or in their sensing.}
\end{figure}

\subsection{Pooling Across Animals}
\begin{figure}[t]
\includegraphics[width=\textwidth]{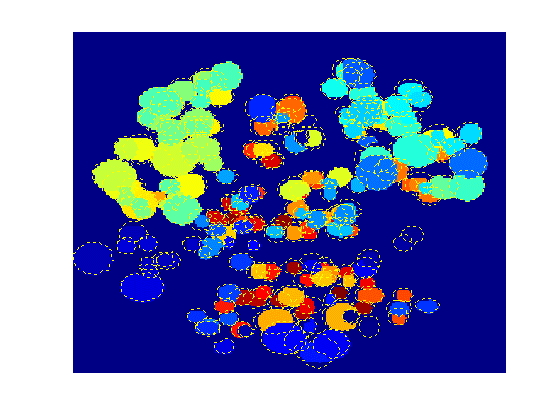}
\caption{Computed canonical ganglion for H. verbana derived from the correspondence matching algorithm. We used the results of the correspondence matching algorithm to generate an average or canonical ganglion by computing mean location and size for each cell that was matched across at least three different animals. The shades of neurons are colored according to the weight determined by an LDA projection that would distinguish between swim and crawl models (brighter color mean higher weight; the colors used were arbitrary).}
\end{figure}

Finally in the third step, the framework reconstructs data corresponding to cells that are missing and remain unobserved in some animals. In particular, if we consider the electrophysiological activity for unobserved cells as latent random variables, then we can infer those latent variables by exploiting the fact that they were observed in other animals. Once we have correspondence information across animals, we can fill in missing electrophysiological data. Formally, we invoke data completion via Probabilistic Principle Component Analysis (PPCA) \citep{roweis,tipping}.  To apply PPCA, we construct a matrix Y, where each row corresponds to a neuron and each column corresponds to the fluorescence intensity in a short time interval. Further, since the correspondences between all the animals are calculated, we can stack the data from all the animals in Y such that the rows are arranged according to the discovered correspondences. (we use -1 to denote absence of data due to missing cells in an animal). The PPCA algorithm recovers the low dimensional structure in the data, and inserts missing data via Expectation Maximization \citep{dempster}. The PPCA algorithm starts with an initialized low-dimensional projection and alternates between the E-step and the M-step. The E-step is where the missing data is estimated by considering statistical relationships in the data. The M-step is where the estimates of the low-dimensional projection are further refined.

Consider the matrix $Y$ (dimensions $c \times n$), which consists of neuronal activity recordings of $c$ cells from all the animals, is constructed using the methodology described in text above. We then first scale all the values in the matrix $Y$ between $0$ and $1$. Letâ€™s denote the low-dimensional representation of the data as matrix $X$ (dimensions $k  \times n$, where $k < c$) and the principal components as  $C$ (dimensions $c  \times k$). The PPCA algorithm first initializes the matrices $X$ and $C$ randomly and then alternates between the following two steps:
\begin{align*}
\mbox{E-step: } & \mbox{Estimate } \hat{Y} = CX\\
\mbox{M-step: } & \mbox{Refine } X_{new}=(C^T C)^{-1} C^T \mbox{ and }C_{new}=\hat{Y}X_{new}^T (X_{new} X_{new}^T )^{-1}\\
\end{align*}
The algorithm converges when the maximum change in any of individual dimensions of estimates Y Ì‚ is less than 0.001. PPCA is guaranteed to converge so that it produces data completion even for neurons that are not observed in some animals.

In our implementation, optimization for Step 1 (see Table 1) is performed via Limited Memory BFGS \citep{liu} routine and energy minimization for the above Equation is performed via iterative variational inference \citep{beal}. There are three parameters that need to be specified in the framework:  $c$ the upper limit on cost of allowed matches, the trade-off parameter between compatibility and relative locality measure, and $k$ the dimensionality of the low dimensional projection in PPCA. These parameters are determined via a cross-validation methodology. The cross validation is performed out by considering the aggregated matrix $Y$, randomly reducing $10$\% of the observed data, and considering the reconstruction error using an $L2$ norm on the removed data. This process is repeated $10$ times and parameters with minimum average reconstruction error are chosen. The search space for parameters $c$ and $\lambda$ lie in log-scale (i.e. $c$ and $\lambda \in [10^{-5}, 10^{-4}, .., 10^5]$), while for $k$ we try in a linear range (i.e. $k \in [1, .., 25]$).

\section{Experiments}
Training data for learning the parameter A was collected by an experimentalist (EPF) who hand-annotated 815 different match pairs  across all the animals. Fig. 1 shows the resulting compatibility measure for these data. Note how the physical properties (such as size, relative location, packet membership) of the most likely matches (highlighted by a white outline) illustrate the quality of the learned function. The matching procedure results in a correspondence map (Fig 2.) matching neurons across the 6 different animals. Once the correspondence map was calculated, it was used to generate a prototypic model of animal by averaging physical as well as functional properties (Fig. 3).

\begin{figure}[t]
\includegraphics[width=\textwidth]{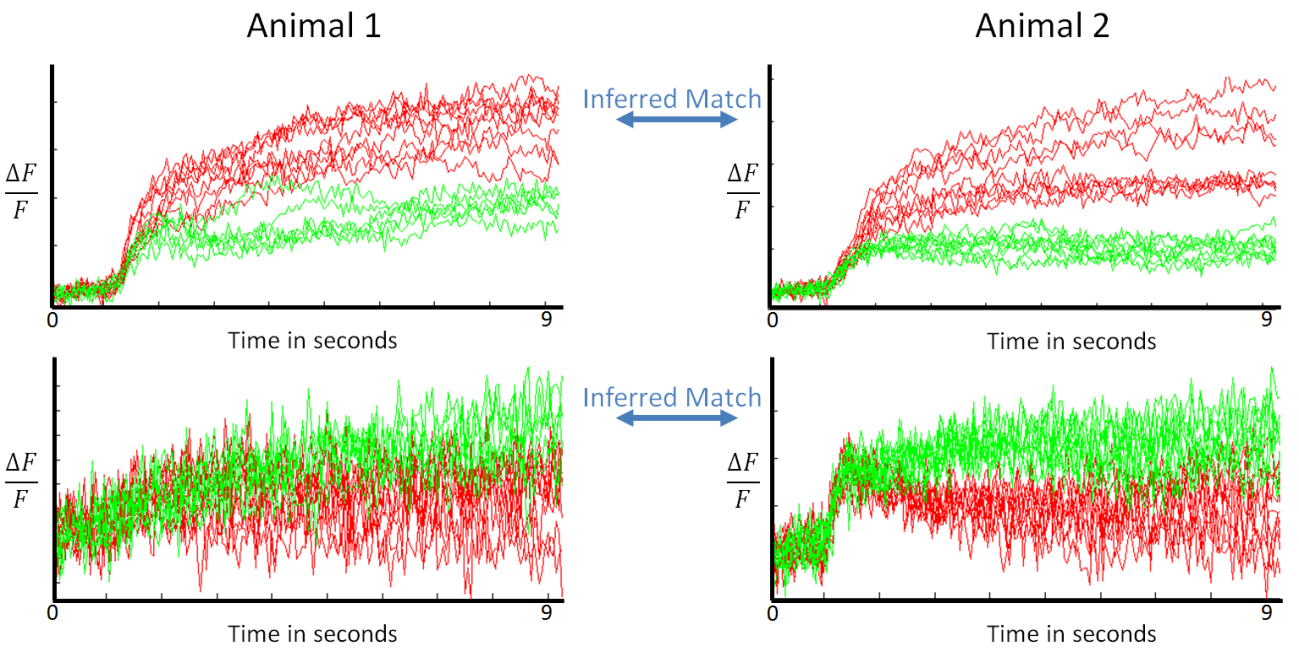}
\caption{Activity of neurons in a leech ganglion from prior study \citep{briggman}, showing how their neuronal activity can be used to identify homologous neurons across animals. Voltage-sensitive dye traces from two different neurons that were considered to be matches by the correspondence matching algorithm. The traces highlight that the algorithm has the capability of recovering correspondence across cell that are functionally similar.}
\end{figure}

Because the resulting correspondence map was computed simultaneously across all the animals, it provides a simple way to analyze the quality of the recovered solution. In addition to the physical properties, the functional characteristics of any two matched neurons across different animals are similar across animals (Figure 4).  We observed that a simple estimator based on average activity of neurons in five animals predicted the activity of the sixth one (Figure 5). Here we estimated the entire time series of activity for a given cell in an animal by considering the activity for the corresponding cell across the rest of the five animals. Two different models for swim and crawl mode are computed where the prediction is performed by computing an average across all the observed time-series.

For all six animals, lower differences between the observed electrophysiological activity and predictions made by a model learned from rest of the animals confirm that the framework had recovered correct correspondences between the neurons across animals.

Although the matching algorithm performs quite well, it is likely that the algorithm is far from perfect. Many of the matched cells may not be correct. Since the functional responses of the cells are a factor for the matching, cells with little functional signal will be harder to match than those with big signals. Cells lacking functional signal, however, are not providing a lot of information for predicting behavioral outcome. Thus, these cells likely have poor matches, but are also likely the cells which are non-relevant to the swim-crawl decision circuit. It is also possible that many cells are effectively the same given this data set, but the matches do not truly reflect homologous pairs. We expect many cells to be functionally the same and mismatching these similar cells may not hurt our analysis.

\begin{figure}[t]
\includegraphics[width=\textwidth]{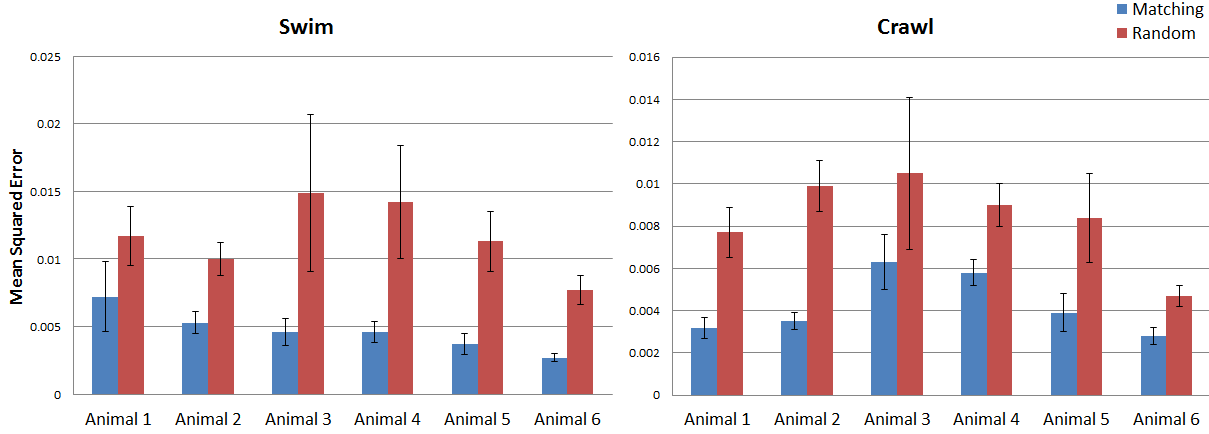}
\caption{Bar graphs that highlight the results to test the recovered correspondence using a leave-one-animal-out analysis. The plots were generated by first considering a candidate test animal and then building a predictive model for each cell (from when the animal swam or crawled) using the remaining five animals. The bar-chart compares mean-squared error between predicted and observed electrophysiological activity when matching using the proposed framework with random selection. The differences across all of the six leave-one-animal-out test cases are significant.}
\end{figure}
\begin{figure}[t]
\includegraphics[width=\textwidth]{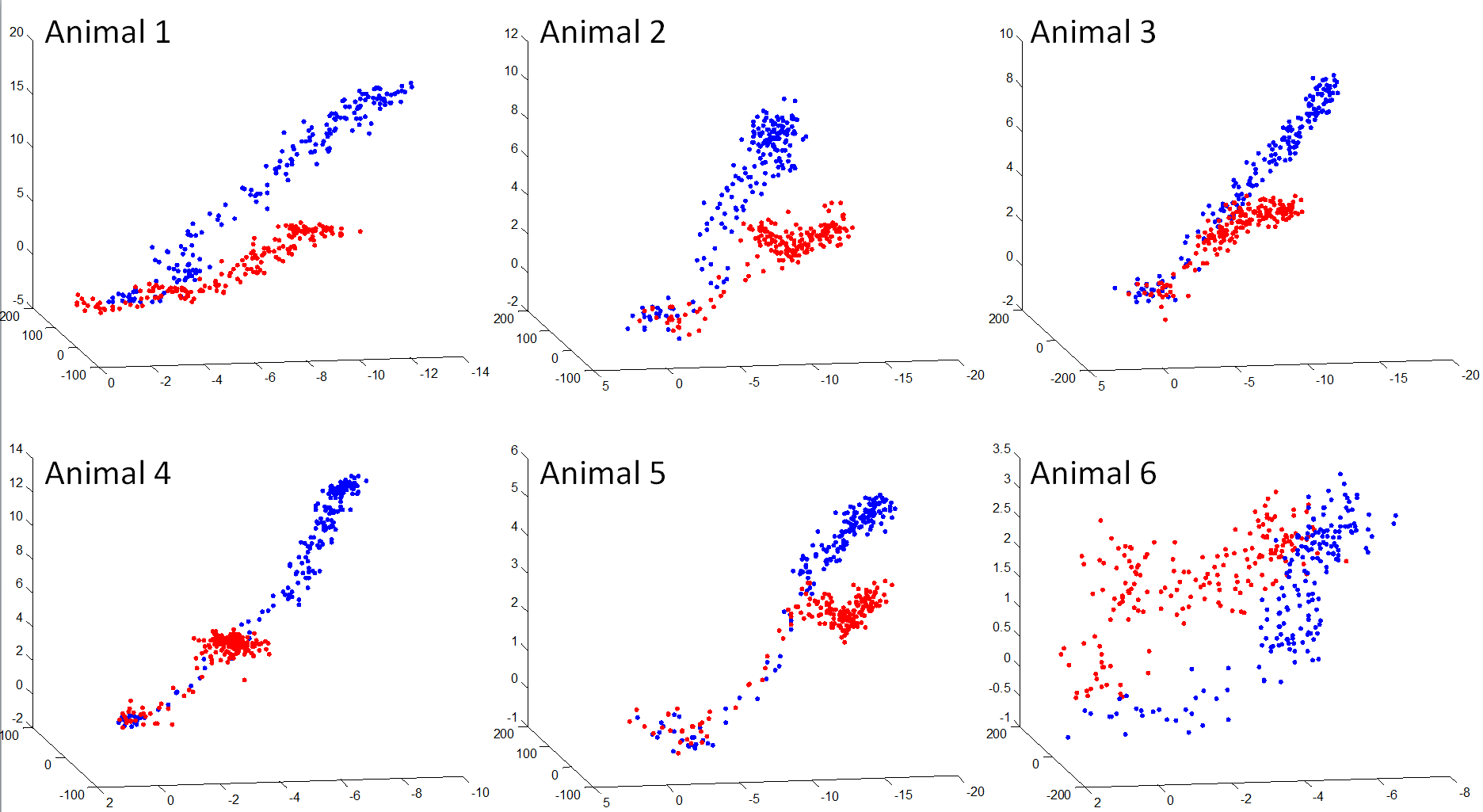}
\caption{Using correspondences to predict behavior from neuronal activity. Identification of corresponding neurons across animals enables larger data sets to be constructed by pooling observations from multiple preparations, which in turn enable deeper and more accurate data analysis to address questions of interest. This figure shows Non-linear projections generated by applying the ISOMAP algorithm. The blue and red dots correspond to swim and crawl mode and depict the trajectory that the voltage-sensitive dye trajectories take for each animal. Note that ISOMAP applied for an individual animal might result in projections that are inconsistent across the different animals. However, using the discovered correspondences of neurons across animals, we combine the data from all six animals, and recover projections that are consistent for all of the animals.}
\end{figure}
\begin{figure}[t]
\includegraphics[width=\textwidth]{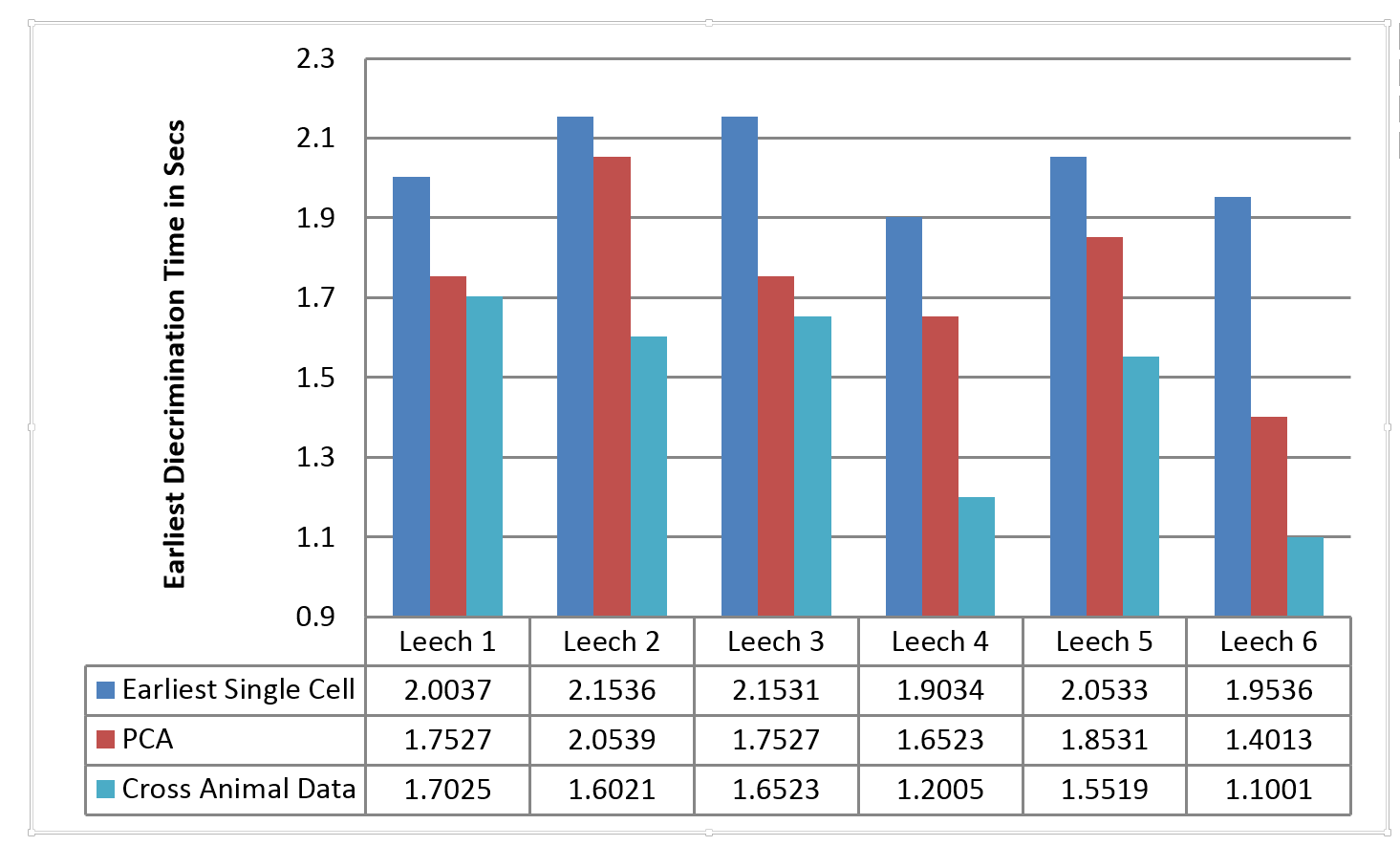}
\caption{Bar graphs showing that pooled data allows us to discriminate between swim and crawl significantly earlier than what was reported earlier using a PCA analysis on data from a single animal \citep{briggman}.}
\end{figure}
\begin{figure}[h]
\includegraphics[width=\textwidth]{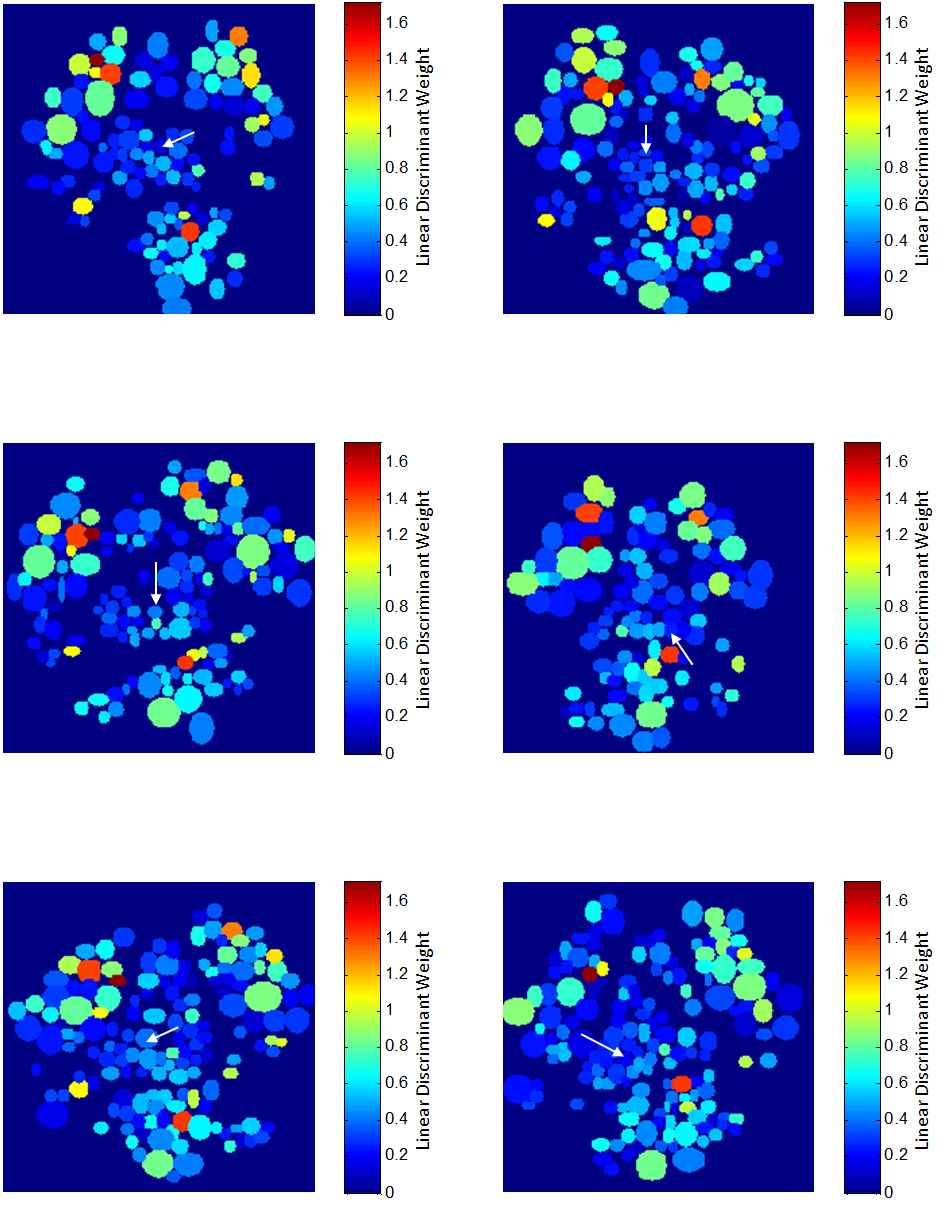}
\caption{Determining influential cells using linear discriminant analysis. The ganglion maps from 6 experiments are shown. The maps are from the same experiments as in Fig. 4. Cells are color-coded based on the magnitude of the contribution to the linear discriminant direction. Red and yellow represent large magnitude contributions, blue represents small contributions. We can see that there are at least 3 cells that are influential and do not include cell 208 (marked using white arrow).}
\end{figure}
The correspondence matching algorithm enables pooling of the data across animals, which allows exploration that was not feasible previously. For example, Figure 6 shows a 3-dimensional projection recovered by applying ISOMAP  \citep{tenenbaum}, a non-linear dimensionality reduction method that is an extension to linear methods such as PCA.  Because the algorithm was applied to the entire pooled data, the recovered dimensions are consistent across all the animals, and thus can be visualized and analyzed within the same reference frames. Previously application of such techniques (such as PCA and LDA in \citep{briggman}) was limited to a single animal at a time resulting in dimensions which were incomparable across animals.

The pooling of data enabled by methodology proved to be valuable in predictive models of decision making. Figure 7 shows that pooling the data across animals enable earlier predictions of one of the two behaviors (swimming or crawling) following stimulation than data from a single animal. Specifically, PCA was performed on pooled data and earliest discrimination time between swim and crawl was determined according the procedure described in \citep{briggman}. In Figure 3, we highlight cells in the composed canonical ganglion that play an important role in the behavioral decisions of the animal.  Combining data across multiple animals enables transfer and overlay of information, allowing aggregation of important statistical parameters and more robust empirical models.  Figure 8 shows ganglion maps for six animals highlighting cells that contribute most towards discrimination amongst the swim and the crawl trials. Note that the highly discriminative cells (towards the red spectrum) are consistent in physical properties such as location and size across the different animals. We also note that these cells are significantly different from cell $208$ that was identified in earlier studies \citep{briggman}.

\section{Related Work}
The work described in this paper builds upon many different sub-areas of machine learning. In particular, the key ingredients include
metric learning, correspondence matching and probabilistic dimensionality reduction \citep{roweis,tipping}.  Distance metric learning is a fairly active research area. Most of the work in distance metric learning focus on $K$-Nearest Neighbor ($k$-NN) classification scenario \citep{duda} and often aim to learn a Mahalanobis metric that is consistent with the training data~\citep{frome, bar, rca, metric-learning}. The distance metric learning method employed in this paper is closest to the work of \cite{roweis_NCA} and \cite{globerson}, but modified to just consider the sets of similar cells given by the user.

Correspondence problems are employed in a multitude of applications. Computer vision is particularly closer to our scenario. Among the simplest are transformations of rigid bodies, where geometry can be exploited \citep{gm99,mcb08}, while correspondences among non-rigid objects, and between non-identical objects, can pose significant challenges. Algorithms applied to more general correspondence problems largely combine the compatibility of points by features with the local geometric compatibility of matches. Such models can be formulated as graphical models \citep{mcb08,tkr08,sh07} or as selecting nodes in an association graph \citep{cll10,css06}, and have been extended to higher-order criteria \citep{dbk09,zs08,lcl11}. Other methods consider the Laplacian constructed from a neighborhood graph \citep{um88,ehl11,mhk08}, and some models are learned from full training examples \citep{tkr08}. Closest to the idea of using reference points are approaches based on seed points \citep{shc11}, landmarks \citep{sj14}, coarse-to-fine strategies \citep{sh07}, and on guessing points that help orient the remaining points in a rigid body \citep{mc12}.

\section{Conclusion and Future Work}
The proposed methodology is likely to be even more useful in combination with other data-centric analyses. For example, the model learned from past data can be employed to guide future experimentation. By computing correspondences between the model and and data from an ongoing experiment in real-time, we can then use the model to guide information extraction strategies.  The methodology can also be extended to perform within-leech analysis, such as discovering bilateral pairs of neurons. In addition, this methodology can readily be used to analyze the simultaneous activity of multiple neurons in other animals. We foresee valuable uses of the approach in overlaying data from larger nervous systems and, moving beyond cells, to higher-level abstractions of nervous system organization, such as components of retina or columns in vertebrate nervous systems. Given its simplicity and the appeal of potentially pooling large quantities of data, the correspondence methodology may find wide use in many areas of neuroscience.

\acks{We acknowledge the assistance of Johnson Apacible, Erick Chastain and Paul Koch.}

\bibliography{sample}

\begin{thebibliography}{39}
\providecommand{\natexlab}[1]{#1}
\providecommand{\url}[1]{\texttt{#1}}
\expandafter\ifx\csname urlstyle\endcsname\relax
  \providecommand{\doi}[1]{doi: #1}\else
  \providecommand{\doi}{doi: \begingroup \urlstyle{rm}\Url}\fi

\bibitem[Bar-Hillel et~al.()Bar-Hillel, Hertz, Shental, and Weinshall]{bar}
A.~Bar-Hillel, T.~Hertz, N.~Shental, and D.~Weinshall.
\newblock Learning a mahalanobis metric from equivalence constraints.
\newblock \emph{Journal of Machine Learning Research}.

\bibitem[Beal(2003)]{beal}
M.~J. Beal.
\newblock Variational algorithms for approximate bayesian inference.
\newblock \emph{University College London}, 2003.

\bibitem[Boykov et~al.(2001)Boykov, Veksler, and Zabih]{boykov}
Y.~Boykov, O.~Veksler, and R.~Zabih.
\newblock Fast approximate energy minimization via graph cuts.
\newblock \emph{IEEE Transactions on Pattern Analysis and Machine
  Intelligence}, 2001.

\bibitem[Briggman et~al.(2005)Briggman, Abarbanel, and Kristan]{briggman}
K.~L. Briggman, H.~D.~I. Abarbanel, and W.~B. Kristan.
\newblock Optical imaging of neuronal populations during decision-making.
\newblock \emph{Science}, 11, 2005.

\bibitem[Bunke(2000)]{bunke}
H.~Bunke.
\newblock Recent developments in graph matching.
\newblock 2000.

\bibitem[Cacciatore et~al.(1999)Cacciatore, Brodfuehrer, Gonzalez, Jiang,
  Adams, Tsien, Jr., and Kleinfeld]{cacciatore}
T.~W. Cacciatore, P.~D. Brodfuehrer, J.~E. Gonzalez, T.~Jiang, S.~R. Adams,
  R.~Y. Tsien, W.~B.~Kristan Jr., and D.~Kleinfeld.
\newblock Identification of neural circuits by imaging coherent electrical
  activity with fret-based dyes.
\newblock \emph{Neuron}, 23, 1999.

\bibitem[Cho et~al.(2010)Cho, Lee, and Lee]{cll10}
Y.~Cho, J.~Lee, and K.~M. Lee.
\newblock Reweighted random walks for graph matching.
\newblock In \emph{European Conference on Computer Vision}, 2010.

\bibitem[Cour et~al.(2006)Cour, Srinivasan, and Shi]{css06}
T.~Cour, P.~Srinivasan, and J.~Shi.
\newblock Balanced graph matching.
\newblock In \emph{Advances in Neural Information Processing Systems}, 2006.

\bibitem[Davis et~al.(2006)Davis, Kulis, Jain, Sra, and
  Dhillon]{metric-learning}
J.~Davis, B.~Kulis, P.~Jain, S.~Sra, and I.~Dhillon.
\newblock Information-theoretic metric learning.
\newblock In \emph{International Conference on Machine Learning}, 2006.

\bibitem[Dempster et~al.(1977)Dempster, Laird, and Rubin]{dempster}
A.~Dempster, N.~Laird, and D.~Rubin.
\newblock Maximum-likelihood from incomplete data via the em algorithm.
\newblock \emph{Journal of Royal Statistical Society Series B}, 39, 1977.

\bibitem[Duchenne et~al.(2009)Duchenne, Bach, Kweon, and Ponce]{dbk09}
O.~Duchenne, F.~Bach, I.~Kweon, and J.~Ponce.
\newblock A tensor-based algorithm for high-order graph matching.
\newblock In \emph{Computer Vision and Pattern Recognition}, 2009.

\bibitem[Duda et~al.(2001)Duda, Hart, and Stork]{duda}
R.~O. Duda, P.~E. Hart, and D.~G. Stork.
\newblock \emph{Pattern Classification}.
\newblock John Wiley and Sons, 2001.

\bibitem[Escolano et~al.(2011)Escolano, Hancock, and Lozano]{ehl11}
F.~Escolano, E.~Hancock, and M.~Lozano.
\newblock Graph matching through entropic manifold alignment.
\newblock In \emph{Computer Vision and Pattern Recognition}, 2011.

\bibitem[Frome et~al.(2007)Frome, Singer, and Malik]{frome}
A.~Frome, Y.~Singer, and J.~Malik.
\newblock Image retrieval and classification using local distance functions.
\newblock In \emph{Advances in Neural Information Processing Systems}, 2007.

\bibitem[Globerson and Roweis(2006)]{globerson}
A.~Globerson and S.~Roweis.
\newblock Metric learning by collapsing classes.
\newblock In \emph{Advances in Neural Information Processing Systems}, 2006.

\bibitem[Goldberger et~al.(2005)Goldberger, Roweis, Hinton, and
  Salakhutdinov]{roweis_NCA}
J.~Goldberger, S.~Roweis, G.~Hinton, and R.~Salakhutdinov.
\newblock Neighbourhood components analysis.
\newblock In \emph{Advances in Neural Information Processing Systems}, 2005.

\bibitem[Gonzalez and Tsien(1995)]{gonzalez}
J.~E. Gonzalez and R.~Y. Tsien.
\newblock Voltage sensing by fluorescence resonance energy transfer in single
  cells.
\newblock \emph{Biophysical Journal}, 1995.

\bibitem[Goodrich and Mitchell(1999)]{gm99}
M.T. Goodrich and J.S.B. Mitchell.
\newblock Approximate geometric pattern matching under rigid motions.
\newblock \emph{IEEE Transactions on Pattern Analysis and Machine
  Intelligence}, 21\penalty0 (4):\penalty0 371–--379, 1999.

\bibitem[Jegelka et~al.(2014)Jegelka, Kapoor, and Horvitz]{sj14}
S.~Jegelka, A.~Kapoor, and E.~Horvitz.
\newblock An interactive approach to solving correspondence problems.
\newblock \emph{International Journal of Computer Vision}, 108, 2014.

\bibitem[Lee et~al.(2011)Lee, Cho, and Lee]{lcl11}
J.~Lee, M.~Cho, and K.~M. Lee.
\newblock Hyper-graph matching via reweighted random walks.
\newblock In \emph{Computer Vision and Pattern Recognition}, 2011.

\bibitem[Liu and Nocedal(1989)]{liu}
D.~C. Liu and J.~Nocedal.
\newblock On the limited memory method for large scale optimization.
\newblock \emph{Mathematical Programming B}, 1989.

\bibitem[Lovasz(1993)]{lovasz}
L.~Lovasz.
\newblock Random walks on graphs: a survey.
\newblock \emph{Combinatorics: Paul Erdos is Eighty}, 2, 1993.

\bibitem[Mateus et~al.(2008)Mateus, Horaud, Knossow, Cuzzolin, and
  Boyer]{mhk08}
D.~Mateus, R.~Horaud, D.~Knossow, F.~Cuzzolin, and E.~Boyer.
\newblock Articulated shape matching using laplacian eigenfunctions and
  unsupervised point registration.
\newblock In \emph{Computer Vision and Pattern Recognition}, 2008.

\bibitem[McAuley and Caetano(2012)]{mc12}
J.~McAuley and T.~Caetano.
\newblock Fast matching of large point sets under occlusion.
\newblock \emph{Pattern recognition}, 45, 2012.

\bibitem[McAuley et~al.(2008)McAuley, Caetano, and Barbosa]{mcb08}
J.J. McAuley, T.S. Caetano, and M.~S. Barbosa.
\newblock Graph rigidity, cyclic belief propagation and point pattern matching.
\newblock \emph{IEEE Transactions on Pattern Analysis and Machine
  Intelligence}, 30\penalty0 (11):\penalty0 2047–2054, 2008.

\bibitem[McKay(1981)]{mckay}
B.~D. McKay.
\newblock Practical graph isomorphism.
\newblock \emph{Congressus Numerantium}, 1981.

\bibitem[Minka(2005)]{minka}
T.~P. Minka.
\newblock Divergence measures and message passing.
\newblock \emph{Microsoft Research Technical Report}, 2005.

\bibitem[Munkres(1957)]{munkres}
J.~Munkres.
\newblock Algorithms for the assignment and transportation problems.
\newblock \emph{Journal of the Society for Industrial and Applied Mathematics},
  5, 1957.

\bibitem[Papadimitriou and Steiglitz(1982)]{papa}
C.~Papadimitriou and K.~Steiglitz.
\newblock \emph{Combinatorial optimization: Algorithms and complexity}.
\newblock Prentice Hall, Englewood Cliffs, NJ, 1982.

\bibitem[Roweis(1998)]{roweis}
S.~Roweis.
\newblock Em algorithms for pca and spca.
\newblock 1998.

\bibitem[Sharma et~al.(2011)Sharma, Horaud, Cech, and Boyer]{shc11}
A.~Sharma, R.~P. Horaud, J.~Cech, and E.~Boyer.
\newblock Topologically-robust 3d shape matching based on diffusion geometry
  and seed growing.
\newblock In \emph{Computer Vision and Pattern Recognition}, 2011.

\bibitem[Starck and Hilton(2007)]{sh07}
J.~Starck and A.~Hilton.
\newblock Correspondence labelling for wide-timeframe free-form surface
  matching.
\newblock In \emph{International Conference on Computer Vision}, 2007.

\bibitem[Tenenbaum et~al.(2000)Tenenbaum, de~Silva, and Langford]{tenenbaum}
J.~B. Tenenbaum, V.~de~Silva, and J.~C. Langford.
\newblock A global geometric framework for nonlinear dimensionality reduction.
\newblock \emph{Science}, 2, 2000.

\bibitem[Tipping and Bishop(1999)]{tipping}
M.~E. Tipping and C.~M. Bishop.
\newblock \emph{Journal of the Royal Statistical Society}, 61, 1999.

\bibitem[Torresani et~al.(2008)Torresani, Kolmogorov, and Rother]{tkr08}
L.~Torresani, V.~Kolmogorov, and C.~Rother.
\newblock Feature correspondence via graph matching: Models and global
  optimization.
\newblock In \emph{European Conference on Computer Vision}, 2008.

\bibitem[Umeyama(1988)]{um88}
S.~Umeyama.
\newblock An eigendecomposition approach to weighted graph matching problems.
\newblock \emph{IEEE Transactions on Pattern Analysis and Machine
  Intelligence}, 10\penalty0 (5), 1988.

\bibitem[Weinberger et~al.(2006)Weinberger, Blitzer, , and Saul]{rca}
K.~Weinberger, J.~Blitzer, , and L.~Saul.
\newblock Distance metric learning for large margin nearest neighbor
  classification.
\newblock In \emph{Advances in Neural Information Processing Systems}, 2006.

\bibitem[Williams et~al.(1997)Williams, Wilson, and Hancock]{williams}
M.~L. Williams, R.~C. Wilson, and E.~R. Hancock.
\newblock Multiple graph matching with bayesian inference.
\newblock \emph{Pattern Recognition Letters}, 18, 1997.

\bibitem[Zass and Shashua(2008)]{zs08}
R.~Zass and A.~Shashua.
\newblock Probabilistic graph and hypergraph matching.
\newblock In \emph{Computer Vision and Pattern Recognition}, 2008.

\end{thebibliography}

\end{document}